\documentclass[epsfile,prc,aps,nofootinbib,showpacs]{revtex4}
\usepackage{graphicx}
\usepackage{epsfig}
\usepackage{amsmath}
\usepackage{amssymb}

\usepackage{color}

\newcommand{\be}{\begin{equation}}
\newcommand{\ee}{\end{equation}}
\newcommand{\bqa}{\begin{eqnarray}}
\newcommand{\eqa}{\end{eqnarray}}

\begin{document}

\date{\today}

\title{ Extraction of  baryonia  from the lightest antiprotonic  atoms}

\author{ B. Loiseau \thanks{e-mail " loiseau@lpnhe.in2p3.fr"}}
\affiliation{Sorbonne Universit\'es, Universit\'e Pierre et Marie Curie, Sorbonne Paris Cit\'e, Universit\'e
 Paris Diderot, et IN2P3-CNRS, UMR 7585, Laboratoire de Physique Nucl\'eaire et de Hautes \'Energies,
4 place Jussieu, 75252 Paris, France}

\author{S. Wycech\thanks{e-mail "wycech@fuw.edu.pl"}}
\affiliation{National Centre for Nuclear Research, Pasteura 7, 02-093 Warsaw, Poland}

\begin{abstract}
Antiprotonic hydrogen and helium  atoms are analyzed.
Level shifts and width are expressed in terms of $\bar{p}$-nucleon  subthreshold scattering lengths and volumes.
Experimental data are compared to  results obtained from the 2009  version of the  Paris $N\bar{N}$ interaction potential.
Comparison with the 1999 version is also made.
Effects of  $N\bar{N}$   quasibound states are discussed.
Atomic 2P hyperfine structure is calculated for antiprotonic deuterium  and the significance of new measurements is indicated.
\end{abstract}


\maketitle

\section{Introduction}

The  lightest hadronic atoms offer a chance to test hadron-nucleon scattering amplitudes just below the thresholds.
This energy region is of special interest in cases of bound or quasibound states in the hadron nucleon systems. Two cases of current interest the  $\bar{p} $ and ${K}^-$   atoms  are  similar in this respect.
Here we discuss  the case of the antiproton.
Nucleon-antinucleon  bound states (baryonia)  have been searched  for since the beginnings of the LEAR era at CERN.
Nothing definite was found, apparently because of two factors.
First, these states are broad because fast annihilation processes and experiments are confronted with heavy backgrounds.
Second, the exclusion principle is not operative  and  a large number of partial  waves may be formed in the $N\bar{N} $  systems.

High selectivity is needed to detect a clear signal.
One  way to reach selected states is a  formation
reaction.
In this way a resonant-like behavior was observed in the decay $ J/\psi \rightarrow \gamma p\bar{p}$.
The experiment performed by the BES Collaboration~\cite{bai03} found an enhancement in  the $\bar{p}p $ close to the threshold energy equal to 1876~MeV.
With a model it was attributed to a subthreshold peak at invariant mass of 1859~MeV.
This was confirmed by the BES~III Collaboration~\cite{abl12}, and the extrapolation  has led  to a  peak at 1832~MeV.
This may be, but not necessarily is,  a confirmation of a  quasibound state.
The valid proof requires a look into the subthreshold region and this also was achieved by the BES Collaboration~\cite{abl05} in the mesonic decays of $J/\psi$ into $\gamma \pi^{+}\pi^{-}\eta'$.
A subthreshold enhancement, the X(1835), was found in this way.

Interpretation, in the $ J/\psi \rightarrow \gamma p\bar{p}$ process, of the close to $\bar{p}p$ threshold enhancement in terms of $N \bar{N}$ potentials~\cite{DED18, SIB05} leads to the conclusion that there exist a broad spin singlet $S$-wave quasibound state.
However, the state obtained~\cite{SIB05}  by Bonn potential is formed in the isospin $I=1$.
That  generated by Paris potential~\cite{PAR09} happens in $I=0$ at the energy   -4.8~MeV with a width of 50~MeV, state which we named ${N \bar{N}}_S (1870)$~\cite{DED09}.
Paris  potential explains the $X(1835)$~MeV enhancement as a result of an interference of this quasibound $S$-wave state with a background amplitude~\cite{DED09}.

Paris 2009 potential generates also a  deeply  bound, isospin 1 $S$-wave state at  about -80~MeV binding.
This state  is far away from the experimentally tested energy regions and may be an artifact of unknown interactions at very short ranges.
Nevertheless,  it exists in the model and we find that after some modifications it may play  a significant role in the  antiprotonic-atomic physics.

There is a class of selective experiments which also allow one to test the  subthreshold energy region.
These are the  measurements of atomic  levels in very light atoms.
The condition  for a direct partial wave analysis is  the resolution of the fine  structure of these levels.
So far, such resolution  was achieved only in the antiprotonic hydrogen~\cite{GOT03, got99} and it is instrumental in fixing  the $N\bar{N}$  potential at low energy~\cite{PAR09}.
Other atoms  useful to study baryonia are antiprotonic deuterium and antiprotonic helium, as, in these atoms, the experimental  data exist.
The advantage of such  systems lies in  their (relatively) simple nuclear structure  and the fact that  $\bar{p}N $  interactions happen on bound nucleons of well defined separation energy.
In this way one can study the $\bar{p}N $  interaction below the threshold, moreover at different subthreshold energies.

In this work we study the  level shifts and widths from $\bar{p}N $ interactions in the  lightest atoms (levels):  $^2$H($2P$), $^3$He(2$P$, 3$D$) and  $^4$He(2$P$, 3$D$).
The number of states is limited to those characterized by very low atomic-nuclear overlaps.
In these  states, multiple scattering expansions converge very quickly and double scattering or interactions on two nucleons contribute small fractions of the level shifts.
Experiments discussed are fairly old,~\cite{ got99,augnp99,sch91}, but have   been analyzed only in a   phenomenological way and never discussed in terms of a well established  $\bar{p}N $ potential.

Understanding of the simplest atoms is important for the  PUMA project at CERN aiming at studies of antiproton capture  from atomic states  formed on unstable nuclei~\cite{PUMA}.
Relative frequency of $\bar{p} p $ and $\bar{p} n $  captures  will be measured with the intention to determine neutron haloes in  unstable nuclei.
These nuclei are likely to contain  loosely bound nucleons, thus kinematic conditions may be similar to antiprotonic-deuterium  and $^3$He atoms.
Knowledge of the   level widths  structure  in these atoms  would be helpful for this project.
Related  experiments  were performed on stable nuclei~\cite{TRZ01}.
In general  the ratio of captures on neutrons and protons is well controlled,  but it was found that some irregularities  happen for loosely bound protons~{\cite{LUB97}.
One hopes that the light  atom studies will allow one to pinpoint the partial wave responsible for these effects and will give constraints  on the parameters describing  heavy antiprotonic  atoms.

\begin{table}[ht]
\centering
\caption{The ratios of  $ N(\bar{p} n) $ and $ N(\bar{p} p) $ capture rates from atomic states.
The second  column  gives  the experimental numbers  obtained in radiochemical experiments~\cite{LUB97}.}
 \label{tab:radio}
  \begin{tabular}{lc}
  \hline\hline\c{}
atom &   $ N(\bar{p} n)/ N(\bar{p} p)$ \\\hline
 $^{96}$Zr &2.6(3) \\
$^{124}$Sn &5.0(6) \\ \hline
$^{106}$Cd &0.5(1) \\
$^{112}$Sn &0.79(14)\\\hline\hline
\end{tabular}
\end{table}
A few selected  ratios of  $ N(\bar{p} n) $ and $ N(\bar{p} p)$ capture  rates  from atomic states  are presented in Table~\ref{tab:radio}.
Two  normal cases  $^{96}$Zr and  $^{124}$Sn indicate neutron haloes.
The other two results $^{106}$Cd  and $^{112}$Sn are  anomalous and point to proton haloes  which cannot be understood by standard nuclear structure models.
Only  part of the effect is related to a sizable differences  ($\sim 3$~MeV) in the proton and neutron separation energies of the valence nucleons in both of these nuclei.
An additional explanation could be related to a fairly narrow $N$-$\bar{N}$-quasibound  state that enhances $\bar{p}$-$p$  absorptions over  $\bar{p}$-$n$  ones in these nuclei.
Existence of a fairly narrow ($\Gamma \leq 10) $~MeV  state  in a $P$ wave was predicted in Paris potential models~\cite{PAR09} and~\cite{PAR99}.
Its position is, however, not well determined by the scattering data  and studies of light atoms might resolve this question.

In conclusion, we argue that there  could well be  two quasibound isospin $ 1 $ $ N \bar{N}$ states,
one in a $S$ wave  and another in a  $P$ wave.
Both are predicted by the recent Paris potential but at  incorrect energies.
An update of this potential model is required for understanding of the planned CERN and GSI proposals concerning  low energy-antiproton research.

This paper is organized as follows.
Section II presents  a formalism used to calculate  the related complex  shifts $ \Delta E - i\Gamma/2 $.
These  are expressed in terms of  averages of  $S$- and $P$-wave $N \bar{N}$ scattering amplitudes derived from the 2009~\cite{PAR09}  and 1999~\cite{PAR99} Paris $N\bar{N}$-interaction potentials and arranged into a sum of multiple scattering series.
At this moment, the fine structures in  these levels,  are, in general,  not resolved and the $N\bar{N} $ amplitudes
have to be  averaged over spin states.
Calculated atomic levels using the 2009 and 1999 Paris-potential models are compared to  data in section III.
This section is also  devoted to the extraction  of baryonium energies and widths as indicated  by the data.
 Finally we calculate the fine structure of $2P$  states in deuterium which may be useful  to pinpoint properties of the baryonia.
 Detailed expressions of the equations used in our calculation are presented in an appendix.

\section{Formalism}

 \subsection{Relation between level shifts and scattering amplitudes} \label{sec2a}

Experiments which  detect the x rays emitted from hadronic atoms provide energy  levels shifted in comparison to the electromagnetic levels by $\Delta E$  from nuclear interactions.
The  level widths $\Gamma$ frpm antiproton annihilation are also provided in this way.
For   a given  main atomic quantum number  $n$  and angular momentum $L$ these complex level shifts are related to the corresponding   hadron-nucleus  $L$-wave scattering parameter  $ A_L$.
The  relation is usually obtained by  expansion in $ A_L/ B^L$ where $ B$ is the Bohr radius.
The scattering amplitudes  are measured experimentally and interpreted in  terms of Coulomb plus  short-ranged potential solutions.
Inner Coulomb corrections are included  in   $A_L$.
For  $S$ waves such a relation,

\begin{equation}
\label{C1}
\Delta E_{nS}   -i \Gamma_{nS}/2 = E_{nS}-\epsilon_{nS}=
 \frac{2\pi}{m_r}|\psi_{n}(0)|^2A_{0}(1+ \lambda_0 A_{0}/B),
\end{equation}
is known as the Deser-Trueman formula~\cite{ DES55,TRU61}.
Here, $\psi_{n}(0)$ is the atomic wave function at the origin and $m_r$ is the antiproton-nucleus reduced mass. Formula (\ref{C1})  is accurate to a second order in $A_{0}/B$, higher terms in this expansion  are not needed  for  small $Z$ nuclei.
It was obtained in the non-relativistic limit. Changes  are to be introduced in the electromagnetic energy $ \epsilon_{nS}$  which is composed of the Bohr's atom energy, $\epsilon_{n}^0=
-m_r(\alpha)^2/2n^2$, corrected for   relativity, radiative effects  and nuclear polarization, $\alpha$ being  the fine structure constant.
In the case of antiprotons, the weak singularity of the Dirac wave function has to be smeared over the nuclear density but this brings no noticeable changes.

The Bohr radius is given by $1/(\alpha\ m_r)$.
In the $1S$ states one has $\lambda_0=3.154$, and with  $ A_0 \approx 1$fm,  the second order term in Eq~({\ref{C1}) constitutes a few percent correction.
Such corrections are negligible in higher angular momentum states.
For these states a simpler linear relation,

\begin{equation}
\label{C2} \Delta E_{nL}-i \Gamma_{nL}/2 =  \epsilon_{n}^{o}
\frac{4}{n} ~\Pi_{i=1}^{L} \left( \frac{1}{i^2}-\frac{1}{n^2} \right
) ~A_{L}/B^{2L+1} ~( 1+ \lambda_L A_L/B^{2L+1}),
\end{equation}
was derived by Lambert~\cite{LAM70}.
The  second order correction  and inner Coulomb corrections for $N$-$\bar{N}$ systems may be  found in Refs.~\cite{CAR92,KLE05}.
For  $2P$  levels $\lambda_1 =1.866 $   and this correction is negligible.
Orders of magnitude of the  shifts are represented by the coefficients   $\Omega_{nL}$ defined  as

\begin{equation}
\label{C2a} \Omega_{nL} =  \epsilon_{n}^{o}
\frac{4}{n} \Pi_{i=1}^{L} ( \frac{1}{i^2}-\frac{1}{n^2})/B^{2L+1},
\end{equation}
which, for the simplest atoms, are given in Table~\ref{tableomega}.
It offers the relation of the  level shifts to the  scattering parameters

\begin{equation}
\label{C3} \Delta E_{nL}-i \Gamma_{nL}/2 =
\Omega_{nL} A_{L}.
\end{equation}

 \begin{table}[ht]
\caption{Bohr radii and $\Omega_{nL}$ coefficients relating the level shifts to the nuclear
scattering parameters  for some light antiprotonic atoms.}
\begin{center}
\begin{tabular}{lcccc}
\hline \hline
 atom    & $ B$[fm] & $ \Omega_{1S}$[keV/ fm] &
$\Omega_{2P}$[meV/fm$^3$]
& $\Omega_{3D}$[$\mu$eV/fm$^5$] \\
\hline
${\bar p}\ ^1$H   & 57.63 & 0.8668  & 24.46  & 0.3591   \\
${\bar p}\ ^2$H   & 43.247 & 1.539  & 77.19  & 2.012 \\
${\bar p}\ ^{3}$He   & 19.22 & 15.59   & 3958   & 522.6    \\
${\bar p}\ ^{4}$He  & 18.02 & 17.77   & 5125   & 770.1   \\\hline\hline
\end{tabular}
\end{center} 
 \label{tableomega}
\end{table}

The measurements of level shifts are  equivalent to the  measurements of  parameters involved in the scattering  amplitudes $f$, which, in the  low energy expansion, are given  in terms of initial  $\boldsymbol{k}$  and final $\boldsymbol{k'}$ c.m. momenta  by

\begin{equation}
\label{flow} f (\boldsymbol{k}, E,  \boldsymbol{k'}) = a_0(E) + 3~ \boldsymbol{k}\cdot\boldsymbol{k'} \ a_1(E).
\end{equation}
In  the scattering on nuclei discussed above we use capital $A_0$ and $A_1$.
In the case of  $\bar{p}N$ scattering we use $a_0$ and $a_1$  which at the threshold energy become the scattering length and volume.
To specify additional quantum numbers  in the $\bar{p}N$  channels  we use notation $^{2I+1\ 2S+1}L_{J}$, where $S,L$ and $J$ are the spin, angular and total momentum of the pair, and $I$ is its isospin.
At this point we remind one that $A_L$ which enter  relations (\ref{C1}) and (\ref{C2})  are from all short-range
interactions  and contain also the inner Coulomb corrections.
With the procedure used here, the Coulomb field is because the target nucleus  and Coulomb effects are
included in the atomic-wave functions.
Hence the basic scattering lengths and volumes  $a_0$ and  $a_1$ for the $\bar{p}$-$N$ systems are calculated without  Coulomb corrections.
The scattering lengths are defined  in the baryon-baryon convention with the negative  absorptive part.
Thus $A _L= Re~ A_L  - i \ \vert Im ~A_L \vert$ and a bound state close to the threshold in the $S$ wave  is signaled by a large \emph{positive}$~Re~A_0. $

To understand  elementary interactions on bound nucleons one needs to know  $a_1(E) $ and $a_0(E)$  in the un-physical region of subthreshold energies.
A  procedure to calculate the relevant extrapolation is given in appendix \ref{off}.

\subsection{Relation  between  level shifts   and subthreshold amplitudes}
 \label{sec2}

 Let us consider the  antiproton,  bound into an atomic orbital,   scattering on a nucleon, bound in  a nucleus with  a separation energy $E_s$.
 Atomic levels  are calculated, here,  in a  quasi-three body system as represented in Fig.~\ref{fig3body}, and consisting of the antiproton,  of the  nucleon and   of a residual nucleus, $R$.
Such a model makes
 sense only for peripheral antiprotons and we  limit ourselves  to such cases.
Three-body Jacobi coordinates are essential and our notation is specified in Appendix \ref{rec}.
We intend to study energy dependence of elementary $\bar{p}N$  scattering parameters  and it is important to know the energy involved  in the  c.m. of the interacting  $\bar{p}N$ pair.
It  is located in  the subthreshold region partly because of the bindings,  nuclear $E_s$,  atomic $E_a$   and partly  to recoil energy,  $E_{r}$, of the $\bar{p}N $  subsystem with respect to the  residual nucleus.
The values and ranges of the   energies in question,

\begin{equation}
\label{Ecm}
E_{cm} = -E_s- E_a - E_r,
 \end{equation}
are indicated in Table~\ref{tablesuben}  for  the simplest first order collisions.
Our discussion is limited to cases where the Born term gives a predominant result.
 The details of the formalism and of the calculation are given  in  appendices~\ref{off} and~\ref{rec}.
Below we present  the main  formulas.

The inspection of Fig.~\ref{fig3body} shows that we have to use two related pairs of Jacobi coordinate systems: $(\boldsymbol{k}_{12}, \boldsymbol{p}_3)$ pair useful to describe antiproton-nucleon interactions  and another $(\boldsymbol{k}_{23}, \boldsymbol{p}_1)$ pair, useful to describe nuclear and atomic wave functions.
Correspondingly, in coordinate representations, one has the $(\boldsymbol{r}_{12}, \boldsymbol{r}_3)$ pair and the  $(\boldsymbol{r}_{23}, \boldsymbol{r}_1)$ one.
The relation between these last two  pairs   is given by
\begin{equation}
\label{Cj1}
\boldsymbol{r}_{23} =  c~\boldsymbol{r}_{12}  - \boldsymbol{r}_3,   ~~ \boldsymbol{r}_{1} = \gamma\ \boldsymbol{r}_{12}  + \beta \boldsymbol{r}_3,
\end{equation}
where
\begin{equation}
\label{Cj2}
 c =\frac{M_{\overline{p}}} { M_{\bar{p}} +M_N}, ~~  \gamma = \frac{ M_N( M_{\bar{p}} +M_N+M_R)}{(M_{\bar{p}} +M_N)(M_N+M_R)}, ~~  \beta =  \frac{M_R }{ M_R +M_N}.
\end{equation}

In this section, to simplify  the notation, we use the  two coordinates $\boldsymbol{r} \equiv \boldsymbol{r}_{12}$ and $ \boldsymbol{\rho} \equiv \boldsymbol{r}_{23} $    that  correspond to the $\bar{p}$-$N$ interaction  range and  to  the  $N$-$R$ internuclear distance, respectively (see Fig.~\ref{fig3body}).

\begin{figure}[ht]
\includegraphics[scale=0.9]{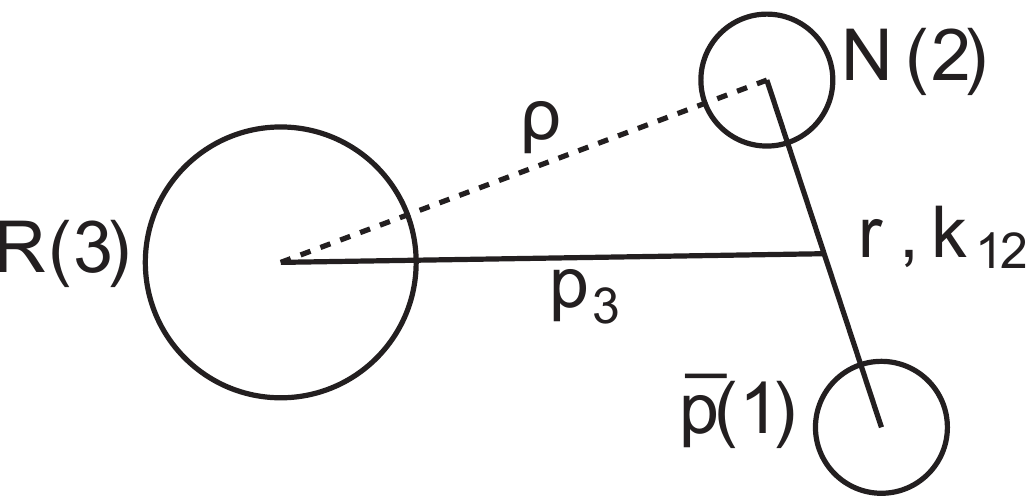}
\caption{Quasi-three-body system: (1) antiproton, (2) nucleon, and (3) residual system. Jacobi coordinates: momentum  $ \boldsymbol{p}_3, \boldsymbol{k}_{12} $  and  space $ \boldsymbol{\rho, r}$.}
\label{fig3body}
\end{figure}

Let us consider the $S$-wave interactions first.
The basic $\bar{p}N $   interaction potential  is

\begin{equation}
\label{C5}
V_{{\bar p} N}(E_{cm},S)=
 \frac {2\pi}{ \mu} \widetilde{T}_0(r,E_{cm})
\end{equation}
where  $ \widetilde{T}_0 (r,E_{cm})  $ is an off-shell scattering matrix given by Eq.~(\ref{soff0}).
Here, isospin indices are not specified, $\mu$  is the reduced mass in the $\bar{p}N $ system.
For a while, we go to  local zero range limit and obtain

\begin{equation}
\label{C6}
V_{{\bar p} N}(E_{cm},S)=  \frac {2\pi}{ \mu} ~a_0( E_{cm})
\delta( \boldsymbol{r})
\end{equation}
where $a_0(E_{cm})$ is the scattering amplitude  at a given c.m. energy,

\begin{equation}
\label{C6a}
a_0( E_{cm}) = \int d\boldsymbol{r}~ \widetilde{T}_0(r,E_{cm}),
\end{equation}
see Eq.~(\ref{soff2})  for further explanations.
Let us notice that potential (\ref{C5}), applied to  $1S$ hydrogen atom reproduces Deser-Trueman  formula~(\ref{C1}) in the leading order, up to inner electromagnetic corrections.

If the antiproton is bound into an atomic orbital,  the energy shifts of \emph{upper} levels  (or levels of small atomic-nucleus overlap) are generated by perturbation  and in the leading order

\begin{equation}
\label{C6b}
 \Delta E_{nL}-i \Gamma_{nL}/2  =  \Sigma_j ~
\langle \psi_L~\varphi |V_{{\bar p} N_j}(E,S)|\varphi~\psi_L \rangle,
\end{equation}
where  the sum over $j$ extends over  all nucleons of the nucleus.
In general, the right-hand side of Eq.~(\ref{C6b}) consists of a complicated $4^3$ dimension integral   which reduces   to easier integrations because of the simplicity of $ V_{{\bar p} N_i}$.
Relevant manipulations in momentum space are explained in Appendix~\ref{rec}.
The main purpose of our formulation is to separate  the factor of atomic-nuclear overlap and the calculation of the length $ a_0 (E_{cm})$ averaged over the recoil energy.
The wave function $\varphi $  of the struck nucleon is  determined in the relative   $N$-$R$ coordinate $\boldsymbol{\rho}$ (see Fig.~\ref{fig3body}).
 The last point allows us to use  the proper asymptotic form of the wave function given by the separation energy.
 It is important  as we discuss interactions localized at nuclear surfaces.
  The Coulomb-atomic-wave functions of given angular momentum $L$,  denoted by  $\psi_L $, are given in terms of antiproton-nucleus coordinates.
 The detailed development of  formula (\ref{C6b}) which, requires  re-couplings of  two basic Jacobi coordinate systems,  is discussed in the appendix.

The recoil energy  is given by $E_r = \boldsymbol{p}^2/2M $  where $\boldsymbol{p}$ is the total momentum of  the $\bar{p}$-$N $ pair and $M=\mu_{R,{\bar p} N}$ is the reduced mass of this pair and   of   the residual nucleus $R$.
As $\boldsymbol{p}$ is not a good quantum number,  formula (\ref{C6b})  involves an integral over $\boldsymbol{p}$.
To obtain an intuitive picture one  has to make calculations in momentum space.
First let us  define an average over recoil energy as

\begin{equation}
\label{C7}
\bar{a}_0 =\int a_0(-E_s- E_a -\frac{\boldsymbol{p}^2}{2M}) ~
\frac{\vert \widetilde{F}_L(\boldsymbol{p}\vert^2 d\boldsymbol{p}}
{\int \vert \widetilde{F}_L(\boldsymbol{q}) \vert^2~ d\boldsymbol{q} }.
\end{equation}
The extent of the recoil energies is determined  by $\widetilde{F}_L(\boldsymbol{p})$  which is calculated in Appendix \ref{rec} and given by  the formula (\ref{r4}).
It turns out to be a Fourier transform  of the atomic and nuclear overlap

\begin{equation}
\label{C8}
 F_L(\boldsymbol{\rho}) =  \varphi(\boldsymbol{\rho}) \psi_L (\beta\boldsymbol{\rho})
\end{equation}
with $\beta$ given in Eq.~(\ref{Cj2}).  It  reflects the fact that the atomic wave function is given in the $R$+$N$  center of mass system and the nuclear wave function depends on the  relative $R$-$N$ coordinate.  The level shifts are  then expressed in terms of  averaged scattering length and an integral over the overlap

\begin{equation}
\label{C9} \Delta E_{nL}-i \Gamma_{nL}/2 =  \frac {2\pi}{ \mu} ~\sum_j ~  (\overline{a}_0)_j ~\int d\boldsymbol{\rho}\ \vert F_L(\boldsymbol{\rho})\vert^2
\end{equation}

\begin{table}[ht]
\caption{The energy regions in MeV entering  the ${\overline{p}}$N amplitudes
involved in the leading interaction terms. 
The first column specifies atomic states of the lightest $ \overline{p} $ atoms. 
The numbers give average energies in the subthreshold region  composed of separation and recoil components.
 The first number in the bracket is the proton separation energy and the second one that of the neutron.
Recoil energies  make no distinctions between protons and neutrons as, apart from the asymptotic region, the wave functions are not that well controlled. 
Differences amount to a small fraction of MeV. 
Numbers in  parentheses indicate approximate widths of the recoil energy distributions. 
  All numbers are rounded off to 0.1 MeV. }
\begin{center}
\begin{tabular}{llll}  \hline\hline
Atom$\backslash$State    & $1S$ & $2P$  & $3D$  \\  \hline
${\overline{p}}$ $^{1}$H   & 0& 0  & 0  \\
${\overline{p}}$ $^{2}$H   &$ -2.2-8.9(5)$&$  -2.2 -5.4(2)  $&$-2.2  -5(1)$   \\
${\overline{p}}$ $^{3}$He   &$ [{-5.5 -7.7}]/{2} -11(7) $& $[{-5.5 -7.7}]/{2}-8.9(2)  $&$ [{-5.5 -7.7}]/{2}-7.3(1)$   \\
${\overline{p}}$ $^{4}$He   &$ {[-19.8-20.6]}/{2} -13.8(10)$&$   {[-19.8-20.6]}/{2}-13.7 (10) $ &$ {[-19.8-20.6]}/{2}- 13.7(10)$   \\\hline\hline
\end{tabular}
\end{center}
\label{tablesuben}
\end{table}

The energies involved in Eq.(\ref{C7}) cover some un-physical subthreshold region.
The  proton and neutron separation energies together with their average, the average recoil energies  and the spread  of recoil energies are given in Table~\ref{tablesuben}.
These values  and the separation energies  indicate that  three atoms $^2$H, $^3$He, $^4$He  cover disjoint   energy regions from about -5 down to -40~MeV.
Therefore, a model is required for the subthreshold  extrapolation.

For $P$-wave-antiproton interactions  we use another pseudo-potential

\begin{equation}
\label{P1}
V_{{\bar p} N}(E,P)=
 \frac {2\pi}
 {\mu}~ 3~ a_1(E_{cm})
    \overleftarrow{\nabla}\delta( \boldsymbol{r})\overrightarrow{\nabla}
\end{equation}
where $a_1(E_{cm})$ is the scattering volume  at a given c.m. energy.
Here, spin indices are not specified.
The calculations of  the recoil energy and overlap integrals are performed in similar manner as for $S$ waves  and details are given in the appendix \ref{rec}.
 Averaging over recoil is  done as in Eq.~(\ref{C7}), the weighting function  being the same.
 Formulas for overlaps are more involved and may also be found  in appendix \ref{rec}.

\section{Results obtained with Paris potential model}

The lowest 12 states in the two  H and He  lightest  elements  are listed in Table~\ref{tableomega} where the appropriate overlap $\Omega_{nL}$
coefficients are presented.
The standard x-ray techniques allow a direct  determination of the  ``lower" line shapes and the
extraction of the ``upper" level width by the intensity loss.
The experimental data  from  deuterium and helium  consists of three level shifts and  five  level widths; see Tables~\ref{tabledeuter},~\ref{table3heI}  and~\ref{table4heI}.
The results from hydrogen have already been used to fix the low energy parameters of the Paris 2009 potential~\cite{PAR09}.  Here we discuss only $2P$  shifts  and $2P, 3D$ widths which can be  described by a corrected first order perturbation.

The Paris potential is a semiphenomenological model used to describe nucleon-antinucleon scattering.
It is based on fitting some 4300~data.
Two versions are discussed, Paris 1999~\cite{PAR99}  (denoted hereafter Paris~99) built to describe $\bar{p} p $ elastic and inelastic scattering data  and an updated version, Paris 2009~\cite{PAR09} (denoted hereafter Paris~09) including also  $\overline{n}p$  scattering  and antiprotonic-hydrogen data.
With respect to subthreshold properties these  two versions differ in the position  of the $^{33}P_1$ quasibound state.
Corresponding energies and widths are $ (E_B,\Gamma/2) = (-4.5,9) $~MeV for Paris~09 and  $ (E_B,\Gamma/2) =~(-17, 6.5)$~MeV for Paris~99.
Also,  Paris~09 predicts an $^{11}S_0 $ quasibound state at  energy $E = -4.8$  and width $ 52 $~MeV; there is no $S$ wave bound state in Paris~99.
Another sizable difference is $ a (^{3}P_2) = 7.22-i12.9$~fm$^3$ for Paris 2009  and $    8.44-i~8.12 $~fm$^3$ for Paris~99.
The preference of  $ a (^{3}P_2)$  cannot be definitely established~\cite{PAR09} by the  hydrogen data~\cite{got99}, which offers a sum of three fine structure lines.

These differences could possibly be detected in terms of the $4P_{3/2}$   component in the deuteron [see below~Eq.~(\ref{s7})].
However, Table~\ref{tablesuben} indicates that fine structure in the $2P$ states in deuterium tests  a narrow energy region around -7.6~MeV below the threshold not too far from the position of $^{33}P_1$ quasibound state predicted by Paris~09.  If this resonance  position is correct it will dominate the  $4P_{3/2}$   component thus  reducing chances to check $ a (^{3}P_2)$.

\begin{table}[ht]
\caption{ $S$-wave scattering lengths in $fm$  and $P$-wave volumes in $fm^3$ for the Paris~99 model.
The subthreshold  energies, -15 and -33 MeV are the central energies in the $\bar p N$ systems in the $2P$ atomic states for $^3$He and $^4$He (seeTable~\ref{tablesuben}).}
\begin{center}
\begin{tabular}{llll}
\hline
Wave                     & ~~~ At   threshold              & At -15 MeV ($^3$He)                & At -33 MeV ($^4$He)         \\  \hline
$^{11}S_0$                     & 0.914114     -i 0.433118   &   2.408820      -i 1.51759    & 3.90277   -i 2.78700      \\
$^{31}S_0$                 &1.116390    -i 0.368901    &    1.767780     -i 0.916545   & 1.77710      -i 1.23273    \\
$^{13}S_1$                 & 1.26410~  -i 0.647898     &   2.075680     -i 1.65776    & 2.71849      -i 2.52060   \\
$^{33}S_1$                & 0.454128 -i 0.355674         &  0.839587  -i 0.542044     & 1.36430     -i 0.674173   \\ \hline
$^{13}P_0$               & ~-8.91649  -i 5.52964         &   -9.49683   -i 3.63118        &-9.84196     -i 2.68574     \\
$^{33}P_0$              & ~~2.73731   -i 9.29717$\times 10^{-04}$  &~2.88113  -i 9.52099$\times 10^{-04}$     & ~2.93514 -i 7.98058$\times 10^{-04}$    \\
$^{11}P_1$              &~-3.85678  -i 0.544166     &  -3.73669     -i 0.415850     &  -3.76583     -i 0.376099 \\
$^{13}P_1$              & ~~5.11843  -i 1.81089$\times 10^{-02}$     & ~5.50724   -i 2.75954$\times 10^{-02}$      &  ~5.63695     -i 3.26629 $\times 10^{-02}$  \\
$^{31}P_1$              &~~1.24447     -i 0.582186   &   1.280490     -i 0.790867   &  ~1.26568      -i 1.04371     \\
$^{33}P_1$              & ~0.199887      -i 1.13695    &~0.361914      -i 6.58659      & -5.74199~    -i 1.51859      \\
$^{13}P_2$             &-0.546355     -i 0.535542    & -0.411518     -i 0.390335     &-0.270804     -i 0.294454   \\
$^{33}P_2$             &  -0.168866  -i 0.141335    & -0.129134     -i 0.146239     &   -0.0741856 -i 0.162846   \\
\hline
\end{tabular}
\end{center}
\label{tablea99}
\end{table}

Let us illustrate the differences  between the two Paris models and the dominant factors determining level shifts in the light atoms. 
Table~\ref{tablea99} presents scattering parameters calculated with Paris~99 at the threshold and compared with the same parameters in the subthreshold energy region. 
The latter were calculated at -15 MeV in the case of $^3$He and at~-33 MeV in the case of $^4$He. 
These are central energies in the antiproton-nucleon systems in $2P$ atomic states as can be seen in Table~\ref{tablesuben}. Analogous results for Paris~09 are given in Table~\ref{tablea09}.

\begin{table}[ht]
\caption{ As in Table~\ref{tablea99} but for the Paris~09 model.}
\begin{center}
\begin{tabular}{llll}
\hline
Wave                     & ~~~ At   threshold              & ~~ $^3$Helium                 & ~~ $^4$Helium          \\  \hline
$^{11}S_0$               &  1.257770    -i 1.17353  & 0.412374      -i 5.08086  &   -3.22081      -i 6.18938  \\
$^{31}S_0$               & 0.751638     -i 0.560066 &  0.926035     -i 0.996672 &   ~1.14442     -i 1.2492   \\
$^{13}S_1$               & 1.191230     -i 0.792671i &  1.520740      -i 1.73793 &  ~1.69286      -i 2.06901  \\
$^{33}S_1$               &0.608209     -i 0.435948&  1.041720    -i 0.809038 &    ~1.60232     -i 1.12518  \\ \hline
$^{13}P_0$               &- 8.78263      -i 4.99065 &   -8.84266  -i 3.40655 &     -8.96530 -i 3.12356     \\
$^{33}P_0$               &  ~~2.73583     -i 0.281606$\times 10^{-02}$ &  ~2.87893 -i 0.317125$\times 10^{-02}$ &    ~2.93932     -i 0.291960$\times 10^{-02}$   \\
$^{11}P_1$               & ~-3.58217     -i 0.320420     &    -3.50523     -i 0.231635 &   -3.50802   -i 0.175309    \\
$^{13}P_1$               & ~~5.12018 -i 0.188749$\times 10^{-01}$      &     ~5.53295 -i 0.297656$\times 10^{-01}$   &    ~5.65938 -i 0.360927$\times 10^{-01}$    ~\\
$^{31}P_1$               &  ~0.997969     -i 0.768087    &   ~0.798880     -i 0.954818  &  ~0.499308  -i 1.00163    \\
$^{33}P_1$               &  ~0.285054      -i 4.107950    &    -5.341040      -i 2.76101  &  -4.192330     -i 0.536913    \\
$^{13}P_2$               & -0.486836     -i 0.870413    &   -0.494158     -i 0.768088  & -0.508217   -i 0.692099    \\
$^{33}P_2$               &  -0.134898  -i 0.210685      &    -0.100130 -i 0.231022     &-0.0588264  -i 0.266901   \\
\hline
\end{tabular}
\end{center}
\label{tablea09}
\end{table}

The atomic  level shifts an widths are given by statistical averages of scattering lengths and scattering volumes. This procedure averages 
strong long distance effects of the  pion exchange force. For example, the tremendous scattering  real volume in   the $^{13}P_0$   state is off set by the real scattering volume in the $^{33}P_0$ which has 3 times higher statistical weight. As a net effect it is the quasibound states that stick-out above the averages. On the other hand it is the spin splitting of atomic states that offers a chance to determine some specific  partial waves. 

The $^{11}S_0$  displays a quasibound state in Paris~09 which is visible as strongly attractive amplitude in the deep subthreshold region met in $^4$He.
This attraction is not supported by the data discussed below. On the other hand there is a strong evidence from BES \cite{abl05} that such a  bound state 
exists. This  controversy  requires  a solution and we will suggest a way out. 

Another resonant amplitude is  $^{33}P_1$  which in Paris~99 is predicted to happen in the energy region tested in $^3$He and for Paris~09 it happens in the 
region of Deuteron. One consequence is that Paris~99 provides a  reasonable description of deuteron shifts and Paris~09  misses the shift badly.

\subsection{Antiprotonic deuterium atoms}

The calculated and experimental  shifts and widths  are  compared  in table~\ref{tabledeuter}.
The results  indicates  $P$-wave dominance and strong sensitivity to the $^{33}P_1$ quasibound state energy, see  Fig.~\ref{figsplit} for a visualization.
Paris~09  solution   yields an unacceptable attractive  level shift and an excessive level width.
That is because of strong overlap of the  quasibound state with the  $E_{cm}$ region characteristic for the deuterium.
It is evident that bound state position of -4.8~MeV  given by the Paris~09 model should be pushed down at least to the $[-10,-8] $~MeV region   which is below  $-7.6 $~MeV characteristic for  $\bar{p}N $  c.m. energy in  the $2P$ atomic state, see Table \ref{tablesuben}. This would reduce overlap of the c.m. energies with the quasibound state  reducing the excessive width of the $2P$ atomic level.
In addition, locating the quasibound state below the  $ \bar{p}N $ energy   will produce  the  missing repulsion to the $2P$ level.
On the other hand  the $^{33}P_1$ quasibound state generated at -17~MeV by Paris~99 model is too far from  the $ \bar{p}N $  c.m. energies and the  absorption width of the $2P$  level becomes too small.
Hence, the experimental atomic levels in deuterium    suggests the proper  position  for the $^{33}P_1$  quasibound state to be located in between the two Paris solutions.
Similar conclusion follows from Table \ref{tab:radio} indicating anomalies of neutron haloes  obtained in the radiochemical measurements.
In the two controversial  nuclei $^{106}$Cd  and $^{112}$Sn  one finds proton separation energy $ S(p)= 7.35$ and $7.55$~MeV and neutron separation energy $ S(n)= 10.87 $ and $10.79 $~MeV, respectively.
In comparison to other tested nuclei, these two, present a sizable difference in the  neutron and proton separation energies.
These differences are comparable to the  $^{33}P_1$  quasibound state width.
Neutron haloes, tested by the radiochemical method, involve extreme nuclear surfaces and the antiproton captures occur predominantly on the valence nucleons.
Characteristic   $\bar{p}n $ energies in their c.m. systems  are about -16~MeV and  corresponding   $\bar{p}p $ c.m. energies are about -11~MeV.
At these c.m. energies  Paris~09 potential  yields  relative  capture rates $ R_{n/p} \simeq 0.9 $ for both $S$ and $P$ waves in the  $\bar{p}N $ systems.
Actually, that value is commonly used in the analyses  of the radiochemical experiments~\cite{LUB97} but it leads to the anomalous proton haloes in $^{106}$Cd and $^{112}$Sn nuclei.
These anomalous results  may be understood if the resonant position is close to the  $\bar{p}p $ c.m. energies and far from the  $\bar{p}n $ c.m. energies.
Such situation may favor $\bar{p}p $ captures imitating  proton nuclear haloes.
To meet  this, the $^{33}P_1$  state should to be located  in  [-11, -9]~MeV  segment.

\begin{table}[ht]
\caption{Results for $2P$-deuterium  level corrections in meV calculated with the spin averaged amplitudes of the Paris~09 potential.
The second and third column numbers are the real and twice the imaginary part of the average complex shift $\Delta$~(\ref{s9}).
Numbers in curly brackets are results with the Paris~99 potential.}
\begin{center}
\begin{tabular}{lcc}
\hline
\hline
order     & Shift        & Width \\  \hline
 $S$ wave & 100\{113\}     & 210\{145\}  \\
$P$ wave  & -9\{58\}       & 365\{206\}   \\\hline
 Sum      & 91\{171\}      & 575\{341\}    \\
Data~\cite{GOT03,got99}      & 243$\pm 26$  & 489$\pm 30$    \\\hline
\hline
\end{tabular}
\end{center}
\label{tabledeuter}
\end{table}

\subsection{Helium atoms}

Experiments measure lower $2P$ and and upper $3D$  atomic levels.
Bulk of the interaction happens in a low density region, although Helium  nuclei have no real surface region. Multiple scattering is  negligible in the $3D$ states  but sizable  in the $2P$ states.
What matters most is a proper single nucleon wave  function.
Here,  we use Eckart functions~\cite{MCC77,LIM73} which have  asymptotic given by separation energies  $E_s$ and a phenomenological form at short distances

\begin{equation}
\label{h1}
 \varphi(r) = [ 1-\exp(-\widetilde{\beta} r ) ]^4~ \frac{ \exp(-\widetilde{\alpha } r)}{r},
 \end{equation}
 where $\widetilde{\alpha}=\sqrt{2\mu_{R,N} E_s}$.
 The short distance parameter $\widetilde{\beta}$ is extracted from electron or pion scattering experiments   and fixed to reproduce  the zeros of  charge form-factors.
 Thus for $^3$He, $\widetilde{\alpha }=~0.45 , \widetilde{\beta}=~1.753 $~\cite{LOC74,FEA77}, and for $^4$He$, \widetilde{\alpha }=~0.846, \widetilde{ \beta }=1.20~$\cite{LIM73}, all in fm$^{-1}$ units.

\begin{table}[ht]
\caption{Leading order calculations  in eV for $2P$ and in   meV for $3D$ (widths only) level corrections in $^3$He obtained  with the spin averaged amplitudes of the Paris~09 potential. Numbers in curly brackets are obtained  with the Paris99 potential.}
\begin{center}
\begin{tabular}{lccc}
\hline
                    &    2P shift                   ~~      &   2P width            ~~  &        3D width\\  \hline
$S$ wave            &      6.68\{9.22\}               ~~     &17.5\{11.5\}           ~~ &        0.69\{0.49\}   \\
 $P$ wave           &     -6.36\{1.44\}               ~~     & 15.0\{26.9\}           ~~  &       1.46\{2.08\} \\
 Sum                &      0.31\{10.71\}                ~~    &32.5\{38.4\}           ~~ &        2.15\{2.57\}  \\
 Data~\cite{sch91}  &      17$\pm 4$                 ~~     & 25$\pm 9$             ~~ &      2.14 $\pm 0.18$  \\\hline
\hline
\end{tabular}
\end{center}
\label{table3heI}
\end{table}

Reference~\cite{WYC85} finds (with a different $N\bar{N}$ potential)   a $2\%$ higher order corrections to the $2P$ levels in deuterium.
Such corrections are of the same magnitude  with the Paris potentials and would not change our conclusions.
On the other  hand in Helium we find sizable multiple scattering corrections.
A method to sum the multiple scattering expansion series is presented in  appendix~\ref{mscat}.
Here we present results.

\begin{table}[ht]
\caption{ As in Table~\ref{table3heI} but only for  $2P$ level  including higher order corrections. Now the contribution
of the $P$ wave interaction depends also on $S$ wave interaction as a result of multiple scattering summation method.}
\begin{center}
\begin{tabular}{lcc}
\hline
                    &    2P shift                   ~~       &   2P width          \\  \hline
$S$ wave            &      6.66\{7.83\}               ~~     &12.3\{7.70\}           \\
 $P$ wave           &     -5.20\{4.75\}               ~~     &19.1\{22.13\}          \\
 Sum                &      1.46\{12.59\}                ~~   & 31.4\{29.8\}           \\
 Data~\cite{sch91}  &      17$\pm 4$                 ~~     & 25$\pm 9$            \\\hline
\hline
\end{tabular}
\end{center}
\label{table3heII}
\end{table}

Tables~\ref{table3heI} and~\ref{table3heII} compare first order and higher order results.
The differences are fairly small and also  difficult to interpret in simple terms.
In general the widths are suppressed which is a standard effect of strong absorption.
Small changes in level shifts reflect interference of first order scattering parameters $ a$   and  second order $a^2 $ which depends on   the signs of real parts.
The average scattering lengths yield repulsive shifts ($ Re~ a_0 > 0$) but  scattering volumes generate attraction  ($ Re~ a_1  <0 $)  below the quasibound state and repulsion ($Re~a_1  > 0 $)  above this state.
Thus the interference pattern of  the single  scattering and double scattering terms  depends on the $\bar{p}N$   c.m. energy.

Paris~99 model is fairly consistent with the Helium data while Paris~09 misses the  repulsion in the $2P$ atomic state.
The downward shift of $^{33}P_1$ required by the deuteron levels does not remove the  inconsistency.  Apparently there is another source of the difficulty and the $^4$He atoms indicate a  new possibility.

\begin{table}[ht]
\caption{As in Table~\ref{table3heI} but for the $^4$He atoms.}
\begin{center}
\begin{tabular}{lccc}
\hline
                    &    2P shift                   ~~      &   2P width            ~~  &        3D width\\  \hline
$S$ wave            &      9.72\{17.6\}               ~~     &26.0\{19.8\}           ~~ &        0.66\{0.50\}   \\
 $P$ wave           &     -9.01\{-10.4\}               ~~    & 14.9\{14.8\}           ~~  &      0.91\{0.91\} \\
 Sum                &      0.708\{7.2\}                ~~    &40.9\{34.6\}           ~~ &      1.57\{1.41\}  \\
 Data~\cite{sch91}  &      18$\pm 2$                 ~~     & 45$\pm 5$             ~~ &      2.36 $\pm 0.10$  \\\hline
\hline
\end{tabular}
\end{center}
\label{table4heI}
\end{table}

\begin{table}[ht]
\caption{ As in Table~\ref{table3heII} but for the $^4$He atoms.}
\begin{center}
\begin{tabular}{lcc}
\hline
                    &    2P shift                     ~~       &   2P width            ~ \\\hline
$S$ wave            &      8.94\{12.3\}               ~~       &  14.7\{10.4\}           ~ \\
 $P$ wave           &     -8.71\{-10.9\}               ~~      & 19.0\{18.6\}          ~ \\
 Sum                &      0.23\{1.4\}                ~~       & 33.7\{29.0\}           ~ \\
 Data~\cite{sch91}  &      18$\pm 2$                  ~~      & 45$\pm 5$             ~ \\\hline
\hline
\end{tabular}
\end{center}
\label{table4heII}
\end{table}
Tables \ref{table4heI} and \ref{table4heII} compare first order and higher order results, respectively.
Both fail to reproduce the repulsive level shifts.
Difficulties of the Paris potential rise with the  increasing distance  from the nucleon-antinucleon threshold.
The   $^4$He atom  presents an extreme case which is not matched even in heavy antiprotonic atoms.
Because of the peripheral nature  of nuclear antiproton captures  the nucleon separation energies as high as 21~MeV are met very rarely.
In some sense  the $^4$He atom  is located on the boundary of unknown  $N\bar{N}$ subthreshold  c.m energies.
Many models of $N\bar{N}$ have predicted deeply  quasibound states which historically have been viewed with scepticism as nothing was determined in direct experiments.  Furthermore  the meson exchange interactions are not very reliable at short distances.
The Paris~09 potential also generates such a deep and broad state at about~-80~MeV in the isospin 1 state. Visualization of the related amplitudes may be found in reference~\cite{HRT18}
 which shows that this state dominates the $S$-wave subthreshold scattering in this deep region.
Now, the difficulty in the understanding of He  atomic data would be solved immediately if the position of this quasibound state is shifted up by some 20~MeV. Both, the missing repulsion and  somewhat weak absorption strength will  reach consistency with the  data.

\subsection{The $R_{n/p}$  ratios}

Different but related  experiments had been performed in the early LEAR era.
These studied captures of stopped antiprotons in deuterium and helium chambers and measured  relative rates  of antiproton captures by  neutrons and
  by protons.
Such ratios are essential for studies of  the neutron haloes in heavier nuclei.
Table \ref{tablernp} gives $R_{n/p}$, the  ratio of  basic ${\bar p}n$ capture rate  to  ${\bar p}p$  capture rate.
It indicates consistency of Paris~09 potential with the $R_{n/p}$ results in $^3$He and $^4$He.
The $R_{n/p}$  was calculated in $2P$ atomic state that is likely to be the state of capture for  $70\%$ of all antiprotons that reach low level atomic  states~\cite{POT77}.
 These  findings  are  also correct if a sizable fraction of antiproton absorption occurs from higher  atomic   $nP$ states.

   On the other hand, the deuterium calculations  given  in table~\ref{tablernp}   have no direct relation to the data  as the capture is likely to happen from many  $nS$ atomic states.
The $nS$  results put  into brackets  are  given by the  Paris potentials, but calculated under the assumption that  captures in $nS$ deuterium levels   happen predominantly  in the  ${\bar p}N$   $S$ waves.
This assumption is based on calculations of Ref.~\cite{WYC85}.
Here, we  are unable to offer the atomic  cascade details  and predict the fraction of  captures from  $nP $ states    relative to  captures from the $nS$ atomic states.
Nevertheless, Table~\ref{tablernp} allows    the conclusion  that Paris~09  sets proper limits on the  $R_{n/p}$    in deuterium too.
Results with Paris~99 are also satisfactory although too high for the ${\bar p}$~$^{3}$He and ${\bar p}$  $^{4}$He atoms.

 \begin{table}[ht]
\caption{The $R_{n/p}$,  ratios.
Second column gives experimental results from antiprotons stopped in bubble chambers.
Third and fourth  columns give the ratios calculated with  the two versions of the Paris potential.
It is assumed that capture occurs from  $nP$ atomic levels.
The experimental  ratio in $^3$He~\cite{BAL87} is given per target $0.35(7)$.
Here it is normalized to values per single nucleon, hence a factor 2.
Results for  captures in deuterons from $nS$ states are given  into angle brackets. }
\begin{center}
\begin{tabular}{lccc}
\hline\hline
 Atom & Experiment   & Paris~09 & Paris~99 \\\hline
${\bar p}\ ^2$H ~ \cite{BIZ74}   & 0.81(3)  & 1.09 $ \langle0.55\rangle$ &0.84  $\langle0.61\rangle$ \\
${\bar p}\ ^2$H ~ \cite{KAL80}   & 0.749(18)  & 1.09 $ \langle0.55\rangle$ &0.84 $ \langle0.61\rangle$  \\
${\bar p}\ ^{3}$He ~\cite{BAL87}  & 0.70(14)  & .65   &1.00   \\
${\bar p}\ ^{4}$He ~\cite{BAL87} & 0.48(3)  & .48   & 0.59  \\ \hline\hline
\end{tabular}
\end{center}
\label{tablernp}
\end{table}

\subsection{Hyperfine structure}

All $N\bar{N}$ potentials are strongly spin dependent.
That generates  hyperfine structure of atomic levels built by strong interactions and superposed on the electromagnetic structure.
Below we present the isospin and spin structure of the basic amplitudes involved in the iso-spin and spin states.

The isospin structure is given by
\begin{equation}
\label{is1}
 A({\bar p} p)  = [a(I=0) + a(I=1)]/2 ;~~~~A({\bar p} n ) = a(I=1).
\end{equation}
For the $1S$ atomic level,  deuteron spin 1 adds to antiproton spin 1/2 to  total  spin doublet and total spin quartet states.
For these, the appropriate amplitudes for  $S$-wave antiproton-deuteron interaction are denoted  $A(\lambda S_{\widetilde {J}}) $  where ${\lambda}=2,4$ denotes doublet or quartet and $\widetilde J$ denotes total angular momentum of the three particles.
The notation for the nucleon-antinucleon pair was defined in Sec.~\ref{sec2a}, viz. $^{2I+1~2S+1}L_{J}$.
The antiproton  interacts on the neutron and proton of the deuteron and we just add both interactions and in the notation used below, in Eqs.~(\ref{s1})$-$(\ref{s8}),  we  suppress the upper isospin subscript and, following Eq.~(\ref{is1}), we define,

 \begin{equation}
 \label{defiso}
 a(^{2S+1}L_{J})= \frac{1}{2}~ a(^{1~2S+1}L_{J})+ \frac{3}{2}~ a(^{3~2S+1}L_{J})
 \end{equation}

One has

\begin{equation}
\label{s1}
 A(2S_{1/2})  = \frac{3}{4}~ a(^1S_0) +\frac{1}{4}~ a(^3S_1) ;~~~~A(4S_{3/2})  = a(^3S_1).
\end{equation}
For $P$-wave interactions in the 1$S$ state one has  expressions  similar to those of Eq.~(\ref{s1}),

 \begin{equation}
\label{s2}
  A(2P_{1/2})  = \frac{3}{4}~ a(^1P_1) +\frac{1}{4}~ a(^3P_1) ;~~~~A(4P_{3/2})  = a(^3P_1).
 \end{equation}
In $2P$ atomic levels, the amplitudes given by Eq.~(\ref{is1}) and~(\ref{defiso}) are mixed  in the total $\widetilde{J }=1/2,3/2,5/2$ three particle states. 
However, in both Paris potentials, the splitting from the $S$-wave interactions is minute and the main effect comes from the $P$-wave antiproton-nucleon interactions.
For the two doublet and  three quartet states in the $2P$ atomic levels, one requires additional terms given by $P$-wave  $N\bar{N}$  amplitudes.
In  doublet  the combinations are

\begin{equation}
\label{s4}
 A(2P_{1/2})  = \left[ 9a(^1P_1) + a(^3P_0)  +2 a(^3P_1)]\right /12,
\end{equation}

\begin{equation}
\label{s5}
 A(2P_{3/2})  = \left[ 9a(^1P_1) +\frac{1}{2}  a(^3P_1)  +\frac{5}{2} a(^3P_2)]\right /12,
\end{equation}
and for the spin quartet

\begin{equation}
\label{s6}
 A(4P_{1/2})  = \frac{2}{3}~  a(^3P_0) + \frac{1}{3}~ a(^3P_1),
 \end{equation}

\begin{equation}
\label{s7}
 A(4P_{3/2})  =\frac{5}{6}~a(^3P_1) +\frac{1}{6}~  a(^3P_2),
 \end{equation}

\begin{equation}
\label{s8}
 A(4P_{5/2})  = a(^3P_2).
 \end{equation}
The hyperfine structure level shifts and widths fullfil the  relation

 \begin{equation}
\label{s9}
\Delta =  \frac{ \sum_{k=1}^{5} ~ ( 2\widetilde{J}_k +1) \Delta_k} { \sum_{k=1}^{5} ~  ( 2\widetilde{J}_k +1)},
\end{equation}
where the complex shifts $\Delta_k$ are given in Table~\ref{tablesplit2p}.
The summation extends over all five states specified in Eqs.~(\ref{s4})$-$(\ref{s8}) and involves both $S$ and $P$ wave interactions. 
We find that the average $\Delta$ differs by less than~8~\% from the shift obtained by the spin averaged $\bar p N$ scattering parameters displayed in Table~\ref{tabledeuter}.
The numbers presented in Table~\ref{tablesplit2p} do not contain electromagnetic splittings which are also of the order of 100~eV~\cite{GOT03}. 

 \begin{table}[ht]
\caption{Hyperfine structure splittings, $\Delta_k= \epsilon_k - i \ \Gamma_k/ 2 $   in eV, predicted for $2P$  states in deuterium.}
\begin{tabular}{lcccccc}
\hline\hline
  &  & Paris 09  &    & Paris 99  &    \\  \hline
$k$&state   & $\epsilon_k$&   $\Gamma_k$  & $\epsilon_k$  &$\Gamma_k$    \\  \hline
1&$2P_{1/2}$            &  - 21& 670    & 204   &  459  \\
2&$2P_{3/2}$            &  33  & 203  &  101   &  323 \\\hline
3&$4P_{5/2}$            & 81  & 282    & 89   &  162   \\
4&$4P_{3/2}$            & 241 & 1250  & 337   &  630   \\
5&$4P_{1/2}$            & 54  & 620    & 79    &  344 \\\hline\hline
\end{tabular}
\label{tablesplit2p}
\end{table}

\begin{figure}[ht]
\includegraphics[scale=0.7]{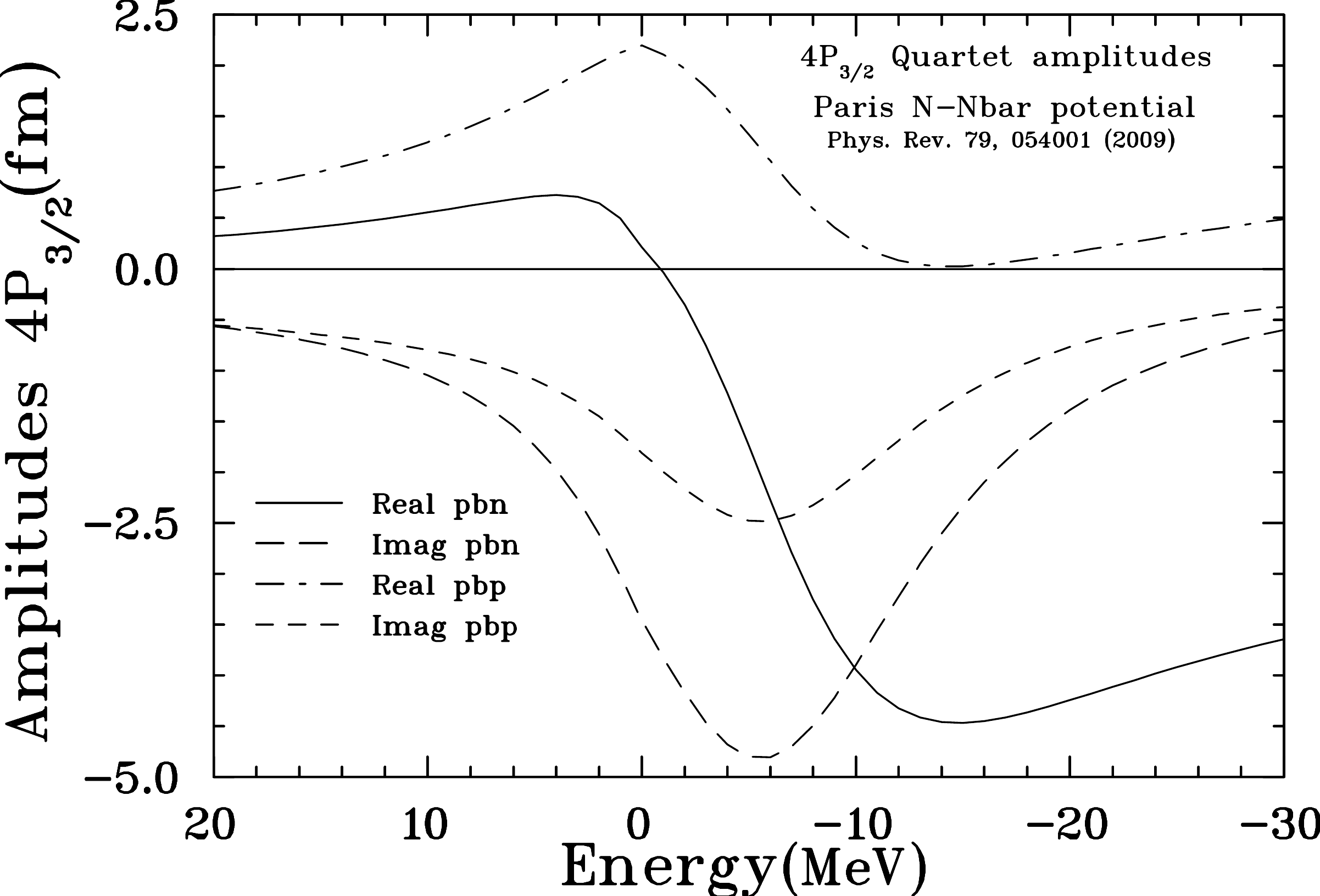}
\caption{Subthreshold amplitudes generating the $4P_{3/2}$ hyperfine structure component in deuterium.   With Paris~09 solution this amplitude is strongly dominated  by the resonant  $ a(^{33}P_1)$  amplitude.   Relevant ${\bar p}N$  c.m. energies fall in the region $-7.6 \pm 1 $~MeV and the resonance position is -4.8~MeV. Downward shift of the resonance position  by a -4~MeV or more will strongly reduce the attraction calculated in this component. In this way the hyperfine structure
splitting is practically nullified. }
\label{figsplit}
\end{figure}

The experiment gives essentially no hyperfine structure splitting but offers  a good check on the widths controlled by direct line structure and  the x-ray intensity loss.
Both Paris potentials generate splitting which is narrow in relation to  the level widths but built  at an incorrect  average level shift (see Table~\ref{tabledeuter}).

One level,  the  $4P_{3/2}$  is singled out and located separately.
Being very broad it can also  be missed in the spectrum.
It reflects strong overlap of  $\bar{p}N$  c.m. energy with the quasibound state spectral density.
As discussed already, a  downward, few MeV shift of the  quasibound state, would  still enlarge the width  and  reduce attraction  removing  practically the fine structure  splitting.
It must be added that this quasibound state shift  will also massively repulse the average shift given in Table~\ref{tabledeuter} as required by the data.
The scenario discussed here is plotted in Fig.~\ref{figsplit}.

On the other hand  Paris~99  gives  a group of five  lines that seem  close to data both in terms of cumulated  widths and  average shift.
In conclusion we see that the data are not consistent with the -4.5~MeV  $^{33}P_1$ quasibound  state but  may tolerate  this state
if its binding is larger.

\section{Conclusions}

Antiprotonic atomic levels  characterized with very small nuclear-atom overlap are a powerful method  to study antiproton-nucleon amplitudes below the antiproton-nucleon threshold.
In this  way one is able to study the structure of the amplitudes in the region reaching down to some -40~MeV  below the threshold.

Comparison of the level shifts and widths in deuterium  and helium atoms with the predictions of Paris potential
model yields following conclusions.

(1) The Paris 2009 potential offers  $S$-wave antiproton-nucleon  amplitudes dominated, just below the threshold,  by a broad quasibound state in $^{11}S_0$ wave.
Interactions in this wave generate strongly repulsive atomic levels in the light atoms.
On the other hand, the $P$-wave interactions are attractive, on average,  unless there  exist  very deeply bound  $P$-wave quasibound states which does not happen in both models.
The balance between $S$-wave repulsion and $P$-wave attraction gives an indication of the uncertain position  of the $^{33}P_1$-quasibound state predicted by two Paris potential models.
 Alas,  the repulsion from the $^{11}S_0$ wave is not strong enough to explain the experimental,  strongly repulsive, levels.
The strongest discrepancy happens in the $^4$He atom where the energies  in  the  $\bar{p}N$ subsystems reach the lower limit of about~-40~MeV.
 The data  require a  strong repulsion which is not given by the model and some enhancement of the absorption strengths predicted by the model.
 All this demands a new phenomenon in the region below~-40~MeV.
 It, in fact, exists in the Paris~09 model and is given by the isospin 1 quasibound state predicted at~-80~MeV.  Consistency with the data requires a shift of this state up to about~-60~MeV.

(2) The level  shifts in light atoms require  some enhancement of nuclear repulsion. This observation is contrary  to the one found in medium and heavy
atoms where attractive antiproton  optical potentials are required.  That happens with  phenomenological potentials~\cite{FRI07} and optical potentials based on  Paris  $\bar{p}N$  amplitudes~\cite{FRI15}. 
The scenario in light and heavy nuclei differ as lower levels in heavier atoms involve higher nuclear  densities and the resulting level shifts are far from the Born approximations. Nuclear many body effects matter. 
It is well understood that the exclusion principle operating in nuclear matter may dissolve  or push up the  $\bar{p}N$ quasibound state.  
This may induce the missing attraction. Reference~\cite{FRI15}  includes this effect  on a Fermi gas model with Fermi momentum $K_{f}\sim \rho^{1/3}$  but finds it not strong enough to bring agreement with the  data. 
A more extreme approach  was studied in Refs.~\cite{GRE82,GRE87} where a different form of $K_f$ was used and a kind of effective self-consistent antiproton nucleon scattering amplitude $T_{eff}$ was calculated.
That may generate required attraction but the price to pay is  a sizable uncertainty in the method used to calculate $T_{eff}$ at the nuclear surfaces. 
It is clear that the problem is not straightforward. In this work we try to check the basic   $\bar{p}N$ subthreshold  amplitudes in a simple situation. 
Studies of heavier atoms and the physics of the involved nuclear surface structure would come next.

(3)  The $P$-wave antiproton-nucleon interaction of Paris potential generates a quasibound state in  $^{33}P_1$ state.
Its position is unstable and  occurs  at $-17$~MeV  in the potential version Paris~99   and at  $- 4.5 $~MeV  in that of  Paris~09.
 Neither solution reproduces  all atomic data.
Consistency with  the $^2$H , $^3$He atomic levels, and understanding of the $R_{n/p}$  anomalies seen in radiochemical experiments  require the state to be located  in the  [-11, -9]~MeV region.

(4) Older data related to bubble chamber measurements of relative $n\overline{p}$ and $p\overline{p}$ capture rates  $R_{n/p}$  are consistent with Paris~09 in all elements. This reflects a fair success of the model in description of the average subthreshold annihilation rates.

(5)   The $^{33}P_1$ quasibound  state dominates the fine structure $4P_{3/2}$  component of $2P$ atomic state in the antiprotonic deuterium.
Unfortunately it is likely to be very broad.
Measurement of the  $4P_{3/2}$ fine structure  would be valuable and would  fix the energy of $^{33}P_1$
quasibound state.

(6) Let us compare the models : on the basis of hydrogen and helium atomic data  it seems that Paris~99 model has some advantage over the later  Paris~09 version.
Definitely the level shifts are better because of the prediction of   the strongly bound $^{33}P_1$   $\overline{N}N  $ state.  
The same state, bound weakly  by the Paris~09 potential, presents difficulties  
 and if this state really exists it should be bound more strongly.
 On the other hand Paris~09 offers some advantages. 
It was based on the addition of anti-neutron  scattering data and the experimental  neutron/proton capture rates $R_{n/p}$ are better reproduced. 
That offers an advantage for  the PUMA project. 
Another advantage of the latter potential is a better description of the BES Collaboration enhancement results~\cite{DED09}.
   
  Both models have problems with the $^{11}S_0$ state which begin already at the level of antiprotonic hydrogen. 
Calculated level shifts are  almost twice as large as the experimental ones~\cite{PAR09}. 
The description of  $\overline{N}N  $ interaction in this state is particularly difficult as the well known pion exchange potential is extremely strong  and may generate a number of quasibound states.  
Various ways to moderate this potential at very short distances lead  to instabilities of the bound state energies.  Now, Paris09 potential offers  a broad but  very weakly bound state  and a deeply bound one.
Probably both states have to be readjusted.
  
To summarize: there is an indication that Paris~09 may be a starting point to offer successful  description of atomic, bubble chamber,  and radiochemical data  provided the position of $P$-wave baryonium is shifted down by about  a few~MeV  and the deeply bound $S$ state is pushed up by some~20 MeV.
That requires an update of this potential model and the work is in progress.

\section* {Acknowledgements}
We wish to thank Jaume Carbonell for helpful discussions   and Detlev Gotta for comments and advice on experimental side.
This study was  initiated by Alexandre Obertelli and PUMA project at CERN.
The  collaboration was supported  by COPIN-IN2P3 agreement No. 05-115.

\appendix
\section {Half-off-shell $\boldsymbol{N\bar{N}}$  interactions \label{off}}

For atomic calculations one needs the off-shell extension of the scattering amplitude in the energy as well as in the momentum variables.
The most general extension for $S$ waves  is given by

\begin{equation}
\label{soff}
T(k,E,k')= \frac{\mu_{N \bar{N}} }{2 \pi  }\int
~\psi_o(r,k)V _{N\bar{N}}(r,E)
\Psi(r,E,k')r^2~dr,
\end{equation}
where $\Psi(r,E,k')$ is the full outgoing wave calculated with the regular free wave $ \psi_o(r,k)= \sin(rk)/(rk)$.
The normalization factor is chosen to produce $T(k,E,k')$ equal to the scattering lengths.
In Eq.~(\ref{soff}) the momentum $k'$ is not related to the energy $E$.
The Fourier-Bessel double transform of $T(k,E,k')$ would generate a nonlocal $\widetilde{T}(r,E, r')$ matrix in the
coordinate representation.
Such involved  calculations  do not seem necessary as the experimental data are not that precise.
We resort to a simpler procedure, standard in nuclear physics,  and for  an application in the antiproton physics see Ref.~\cite{GRE82}.
The subthreshold scattering amplitudes are calculated in terms of an effective $\widetilde{T}(r,E)$
matrix defined in the coordinate representation by

\begin{equation}
\label{soff0}
\widetilde{T}(r,E)= \frac{\mu_{N \bar{N}} }{2 \pi  } ~
V_{N\bar{N}}(r,E)~ \frac{\Psi(r,E,k'(E)) }{\psi_o(r,k'(E))},
\end{equation}
with $k'(E)=\sqrt{2\mu_{N\bar{N}} E}$. In this equation $ \Psi(r,E,k'(E))$ is the solution of the Lippman-Schwinger equation $\Psi~=~\psi_o~+~G^+~V~\Psi$ or of an equivalent Schr\"odinger equation.
To avoid confusion in the notation  we specify the  $\Psi$  in the simplest $S$  wave state. It is
the radial scattering solution  regular at the origin which  behaves asymptotically as

\begin{equation}
\label{soff0as}
\Psi(r,E,k'(E)) \sim \frac {i}{2k'r}[  e^{-ik'r}- S e^{ik'r} ]
\end{equation}
where $S=\exp(i2\delta)$ is the scattering matrix.
The $ \widetilde{T}(r,E)$ is a local equivalent of the nonlocal T matrix in the sense that its  matrix elements fulfill the relation (obtained from Eqs.~(\ref{soff}) and (\ref{soff0} with $k'$ replaced by $k'(E)$) $  T(k,E,k'(E)) = \int dr\ r^2\ \psi_o(r,k) \widetilde{T}(r,E)\psi_o(r,k'(E))$ valid in a narrow subthreshold region where the last integral is convergent.
For positive energies this equation is not practical because of zeros in the denominator of  Eq.~(\ref{soff0}) which occur  at multiplicities of $k'= \pi/ r$. 
One could nevertheless use it for $k'< \pi/ r_{max} $ where  $ r_{max}$ is the distance at which the potential is cut-off.
In this narrow region one finds

\begin{equation}
\label{soff1}
T(k,E,k')= \int~\psi_o(r,k)\ \widetilde{T}(r,E)~\psi_o(r,E,k')r^2~dr,
\end{equation}
thus, in this  case, the $\widetilde{T}(r,E)$  reproduces the half- and full-off-shell amplitudes, allowing to obtain a numerical continuity in calculations.
One needs  also  the Fourier transform of $\widetilde{T}(r,E)$,

\begin{equation}
\label{soff2a}
T(\kappa,E) = \int  d
\boldsymbol{r}~ \widetilde{T}(r,E)~\frac{sin(\kappa r)}{\kappa r},
\end{equation}
and the  relevant $S$-wave   scattering amplitudes  is given by

\begin{equation}
\label{soff2} a(E)=  T(0,E)= \int  d \boldsymbol{r} ~  \widetilde{T}(r,E).
\end{equation}

The main step is to use equation (\ref{soff0}) for  negative energies with  $V_{N\bar{N}}(r,E) $  given by the  Paris  model.
It is used in the Schr\"odinger equation to calculate $ \Psi(r,E,k'(E))$.
Formula (\ref{soff0}) requires numerical care and cutting the  potential tail at very large distances is recommended.
This point, related to the question of relatively  long ranged pion-exchange forces  particularly strong in  the nucleon-antinucleon systems, is discussed below.

\section{The  $N\bar{N}$ force range effects \label{rangeff}}

With no knowledge of the  full off-shell $T(r',E,r)$ matrix, one cannot utilize the  power of Faddeev 3-body equations.
On the level of the pseudopotential method, used here, the  question  is: what is a better description, either  the local zero-range potential of Eq.~(\ref{C6}) or a type of ``folded" potential
 given by  Eq.~(\ref{C5}), with the range given by the half-off shell $\widetilde{T}(E,r)$.
We tried both with the result that the folded potential with the range involved within $\widetilde{T}(E,r)$ is not acceptable.
In particular the range in the absorptive part in $\widetilde{T}(E,r)$ is fairly long from joint impact of short range effects and long range pion exchange. The mean square radius of this range defined by

\begin{equation}
\label{soff3} R_{ms}  = \sqrt{\frac{~{\rm Im} \left[ \int  d \boldsymbol{r} ~  \widetilde{T}(r,E)~ r^2 \right]}{ ~{\rm Im} \left[\int  d \boldsymbol{r} ~  \widetilde{T}(r,E)\right]}}
\end{equation}
amounts to about $1.3$~fm , in both models,  while the range of the basic nucleon-antinucleon potential is about  $0.9$~fm.
In helium  atoms, for $2P$ states, the range effects are moderate and  both  Eqs.~(\ref{C6}) and~(\ref{C5}) yield comparable results.
However,  the range involved in $\widetilde{T}(E,r)$ makes tremendous difference in the upper $3D$ level widths given predominantly by $R_{ms}^4$ .
In those levels, the widths  depend on  the well known asymptotic of nuclear wave functions and are given essentially by the  first order scattering terms.
The experimental data rules out folded potential based on the $\widetilde{T}(r,E) $ by several standard deviations.

A similar result was found  in Refs.~\cite{GRE82, GRE87} where a simple two term separable potential for the $N$-$\bar{N}$  interactions is studied  with some form-factors $v(r)$  representing effects of pion exchange.
The full off-shell $T$ matrix involves terms $ T(r',E,r) \sim v(r')t(E)v(r)$
 where the $t(E)$ matrix is based on the Dover-Richard $N\bar N$ potential. Such terms
in the few-nucleon or nuclear systems yield effective  interaction range,  determined by $v(r)^2$,   roughly  one half of the range given by the half off-shell $\widetilde{T}(r,E)$.

We conclude that localized potential (\ref{C6}),  with range effect hidden in $a(E_{cm})$, is more realistic in the overall approach and it was used in the calculations.
However, with future $N$-$\bar{N}$ potentials,  of  properly   fixed quasibound states positions,  it is advisable to use fully off-shell input which will adapt itself better to states of different atomic angular momenta.

\section {Averaging over recoil energy }
\label{rec}

 Following Sec.~\ref{sec2} we consider the 3-body system depicted in Fig.~\ref{fig3body} and denote the antiproton, the struck nucleon and the residual nucleus as particles 1, 2 and 3, respectively. 
 We adopt the Jacobi coordinates $\boldsymbol{k}_{ij}, \boldsymbol{p}_m $ ($i, j, m=1, 2, 3$, $i< j$) and
in coordinate  representation we use $\boldsymbol{r}_{ij}, \boldsymbol{r}_m  $, but to simplify the notation, we specify only the two coordinates  $\boldsymbol{r}$ and  $\boldsymbol{\rho}$ shown in the figure.

For $S$-wave zero-range interactions the relevant antiproton-nucleon operator in momentum space is

\begin{equation}
\label{r1}
 \widehat{a}_0 =  a_0\left (E_B- \frac{\boldsymbol{p}_3^2}{ 2 \mu_{12,3} }\right)  \delta(\boldsymbol{p}_3-\boldsymbol{p}_3'),
\end{equation}
 where $\mu_{12,3} $ is the reduced mass of the 12 pair and particle 3.
With the wave functions $\widetilde{\varphi}(\boldsymbol{k}_{23})$  and $\widetilde{\psi}(\boldsymbol{p}_1)$  one has the first order expectation value

\begin{equation}
\label{r2}
\langle \widehat{a}_0\rangle = \int  d \boldsymbol{k}_{12}   d \boldsymbol{p}_{3} ~ d \boldsymbol{k}_{12}'   d \boldsymbol{p}_{3}' \widetilde{\varphi}(\boldsymbol{k}_{23}) \widetilde{\psi}(\boldsymbol{p}_1) ~   a_0\left(E_B- \frac{\boldsymbol{p}_3^2}{ 2 \mu_{12,3} }  \right) \delta(\boldsymbol{p}_3-\boldsymbol{p}_3')  ~ \widetilde{\varphi}^*(\boldsymbol{k}_{23}') \widetilde{\psi}^*(\boldsymbol{p}_1').
\end{equation}
To exploit  the $\delta(\boldsymbol{p}_3-\boldsymbol{p}_3') $ function, we use another pair of Jacobi coordinates

\begin{equation}
\label{r3}
\boldsymbol{p}_1 = -\boldsymbol{k}_{12}  - c~ \boldsymbol{p}_3   ;  ~~~~   \boldsymbol{k}_{23} = -\beta~\boldsymbol{k}_{12}+ b\  \boldsymbol{p}_3,
\end{equation}
 where  $ b = 1- c \beta$.

Now we define  a function   $ \widetilde{F}$

\begin{equation}
\label{r4}
\widetilde{ F(}\boldsymbol{p}_3) = \int  d \boldsymbol{k}_{12}  \widetilde{\varphi}(-\beta~ \boldsymbol{k}_{12}+ b~ \boldsymbol{p}_3)  \widetilde{\psi}(-\boldsymbol{k}_{12}  -c~ \boldsymbol{p}_3  ),
 \end{equation}
which allows to present expectation value $\langle \widehat{a}_0\rangle$  as

\begin{equation}
\label{r5}
\langle \widehat{a}_0\rangle =    \overline{a}_0 ~ \int  d \boldsymbol{p}_{3} ~  \vert \widetilde{F}(\boldsymbol{p}_3)\vert^2
 \end{equation}
where
\begin{equation}
\label{r6}
\overline{a}_0 =\int  d \boldsymbol{p}_{3} ~ a_0\left (E_B- \frac{\boldsymbol{p}_3^2}{ 2 \mu_{12,3} }  \right)
 |\widetilde{F}(\boldsymbol{p}_3)|^2   ~ /~ \int  d \boldsymbol{p}_{3} ~  |\widetilde{F}(\boldsymbol{p}_3)|^2.
  \end{equation}
This corresponds to subthreshold value of the scattering length averaged  over some region of recoil energies.  The  last expression   becomes more intuitive in the  coordinate representation.
One obtains Fourier transform of $\widetilde{F}(\boldsymbol{p}_3)$  expressed by

\begin{equation}
\label{r7}
F (\boldsymbol{\rho)} =   \varphi(\boldsymbol{\rho}) \psi(a~\boldsymbol{\rho}),
\end{equation}
i.e. it is an atomic nucleus overlap.
Its Fourier transform determines the  distribution of  the $ N\bar{N}$-pair momentum relative to the residual nucleus.

For $ P$-wave antiproton-nucleon interactions the relevant operator in momentum space is

 \begin{equation}
\label{r8}
 \widehat{a}_1 =  a_1\left(E_B- \frac{\boldsymbol{p}_3^2}{ 2 \mu_{12,3} }\right)  \delta(\boldsymbol{p}_3-\boldsymbol{p}_3') \ \boldsymbol{k}_{12}\cdot \boldsymbol{k'}_{12}.
\end{equation}
The steps from Eq.~(\ref{r2}) to  Eq.~(\ref{r7})  may be repeated.
As before the operator $\widehat{a}_1 $ involves only internal coordinates in the $N\bar{N}$ subsystem and the
average   over recoil is given again by formulas similar to Eqs.~(\ref{r4}), (\ref{r5}) and (\ref{r6}). 
The overlap formula is more involved and in the coordinate representation one obtains (see the definition of $\gamma$ in Sec.~\ref{sec2})

\begin{equation}
\label{r9}
\langle \widehat{a}_1\rangle   =  \overline{a}_1 \Theta \equiv  \overline{a}_1  \int d\boldsymbol{\rho}  ~ \boldsymbol{\partial}_{r}[  \varphi(- \boldsymbol{r}_{}/2 -\boldsymbol{\rho})  \psi(\gamma \boldsymbol{r}_{}  - \beta\boldsymbol{\rho}) ]
|_{\boldsymbol{r}_{}=0}~ \cdot ~
\boldsymbol{\partial}_{r}[  \varphi^*(- \boldsymbol{r}_{}/2 -\boldsymbol{\rho})  \psi^*(\gamma \boldsymbol{r}_{}  - \beta \boldsymbol{\rho} )]
| _{\boldsymbol{r}_{}=0}.
\end{equation}
The overlap may be calculated in spherical coordinates using radial and transversal derivatives $ \partial_r , \partial_T $. For $S$- wave  nucleons only radial derivatives apply, for atomic wave functions  one needs to calculate both derivatives.   
Mixed  atomic-nuclear derivative contributes only to the radial term.

Atomic contribution obtained with $\psi (\boldsymbol{\rho})  = N(n,L)   Y ^m_L  \rho^L  \exp(-\rho /nB) $ becomes

\begin{equation}
\label{r10}
\Theta_A = \int  \frac{d\boldsymbol{\rho}}{4\pi} (\gamma)^2 \left[ \frac{L+1}{2L+1}(\frac{1}{nB})^2  + \frac{L+1}{2L+1} ( \frac{2L+1}{\rho} -\frac{1}{nB} )^2 \right] ~ \left[\frac{ d\psi_r(\beta \rho)}{d\beta\rho}\right]^2 ~ \varphi_r(\boldsymbol{\rho})^2
 \end{equation}

\begin{equation}
\label{r11}
\Theta_N =  \int  \frac{d\boldsymbol{\rho}}{4\pi} ~ \frac{1}{4}~\left ( \frac {d \varphi_r(\boldsymbol{\rho})}{d\rho} \right)^2
\end{equation}
 where $ \psi_r(\beta \rho)$ and  $\varphi_r(\boldsymbol{\rho})$ are  radial parts of the atomic wave  functions ,   $\Theta \simeq  \Theta_A +\Theta_N $  while the mixed term is small.

\section {Summation of multiple-scattering series \label{mscat}}

The perturbation expansion for the level shift  is, up to Coulomb corrections, equivalent to multiple scattering
expansion.
This follows from Trueman formula [see Eqs.~(\ref{C1}) and~(\ref{C2})].
The method of summation used here  is based on simple  formula for meson scattering on two nucleons fixed at a distance $ R$  obtained by Brueckner~\cite{BRU53}.
The scattering length on such  a pair  becomes

\begin{equation}
\label{fm1}
 A =  \frac{ 2a}{ 1+ 2a/ R} .
\end{equation}

The term  $2a / R $ in the denominator  sums  the multiple scattering series.
It has very simple interpretation, 2 is the number of scatterers, $a$ is the scattering amplitude on a single nucleon  and $1/R$   is the propagator for the meson bouncing off the fixed nucleons of the scattered particle.
Expansion of the expression~(\ref{fm1})  in terms of  $2a/R$  gives the multiple scattering series which  is a geometric series.
Equation~(\ref{fm1})  is correct  also for large $2a/R$ when the multiple scattering expansion is divergent.
This simple result indicates a possibility to sum  the multiple scattering series by comparing  it to a geometric  series and correcting the sum at every step of the expansion.
Let us present it as a series for the solution of a pseudo potential built as a multiple-scattering expansion with a potential given by Eq.~(\ref{C6}).
We  denote single scattering term by

\begin{equation}
\label{fm2}
   \langle V \rangle = \langle\psi~\varphi | \Sigma_i   \frac {2\pi}{ \mu} ~ a_i |\varphi~\psi\rangle.
\end{equation}
The double scattering  term involves $G$  a  propagator of the whole system  in-between the subsequent collisions

\begin{equation}
\label{fm3}
   \langle VGV\rangle =
 \langle \psi~\varphi | [\Sigma_i   \frac {2\pi}{ \mu}  a_i]  ~ G~  [\Sigma_j   \frac {2\pi}{ \mu}  a_j] |\varphi~\psi \rangle
\end{equation}
and similarly for higher order terms.
The first order sum is

\begin{equation}
\label{fm4}
\langle V\rangle'  = \frac{\langle V\rangle}{ 1 - \frac{ \langle VGV\rangle}{\langle V\rangle} }.
\end{equation}
Expanding the denominator into a geometric series yields the first two terms of the  multiple scattering.
To have  exact  higher orders, one corrects the denominator.
In this  way one obtains a series in the denominator.
Up to third order

\begin{equation}
\label{fm5}
\langle V\rangle ''  = \frac{\langle V\rangle}{ 1 + C_1 +C_2 +C_3 +..  }
\end{equation}
where  the expansion coefficients are $ C_1 = -\langle VGV\rangle /\langle V\rangle$  , $C_2 = \langle VGV\rangle^2 - \langle VGVG V\rangle /\langle V\rangle  $ and
\begin{equation}
\label{fm6}
C_3 = \frac{\langle VGV\rangle^3} {\langle V\rangle^3 } + \frac{2 \langle VGVGV\rangle\langle VGV\rangle }{ \langle V\rangle^2} -\frac{\langle VGVGVGV\rangle}{\langle V\rangle }.
\end{equation}
For illustration,  we present  the case of an $ S$-wave projectile scattering on  $N$ nucleons  bound in $S$ waves in the nuclear c.m. system.
The wave function is a product of  Gaussian wave functions  yielding  a  density  radius mean square $R$.
At zero projectile energy the propagator is taken to be

\begin{equation}
\label{fm7}
G = -\frac{2m} {4\pi |r-r'|}\ |\varphi\rangle \langle \varphi|
\end{equation}
where $m$ is the reduced mass of the projectile  and of the nucleus, while  $|\varphi\rangle$ is the nuclear wave function.
The terms of expansion may be calculated in an analytic way, and the expansion parameter in this case becomes

\begin{equation}
\label{fm8}
x = \frac{N~a~m}{R \mu},
\end{equation}
 where $\mu$ is the reduced mass of the projectile and of a nucleon of the nucleus.
One obtains $C_1=- 0.977~x$, $C_2= -0.045~x^2$ and $C_3 = 0.0076~x^3$ .
For light nuclei,  like helium,  the series converges rapidly.
For the $2P$  states, this  convergence is much faster, the second order formula~(\ref{fm4}) is sufficient  for the present experimental precision.

Formula~(\ref{fm4}) was checked against a full three-body calculation for low energy $\eta$-deuteron scattering~\cite{DEL00}.
It gives precision of a few $\%$  in a  very demanding case of the virtual  $S$ state close to threshold.
As the deuteron is a loosely bound object one needs a  better propagator allowing also for the deuteron excitation to continuum.
Helium  is  a much stronger bound system and for the $2P$ states such corrections are not necessary.
What is necessary is a better description of the  nuclear wave function $\varphi$.
 This is achieved here by the three-body model of interaction and the proper asymptotic  form of the wave function~(\ref{h1}).
 In this model the  multiple scattering is a series of $ V G V_R +...$ where  $V_R$  is the potential for the antiproton interaction with the residual system.
 For $S$ waves we take it as

  \begin{equation}
\label{fm9}
   V_R =   \frac {2\pi}{ \mu}
 \Sigma_i^{N-1}  ~a_i~\delta( \boldsymbol{r}_{\bar{p}} - \boldsymbol{r}_i).
\end{equation}
 Thus, in this approach, the second order scattering term becomes
  \begin{equation}
 \label{fm10}
  \langle VGV\rangle =
  -\frac{2m} {4\pi} \left[\frac {2\pi}{\mu}\right]^2  \Sigma_{j}^{N}  \Sigma_i^{N-1}   ~a_i ~a_j~ \int\int d\boldsymbol{r}d\boldsymbol{r}' ~{\psi}(\boldsymbol{r}_i) ~ \frac{\varphi(r_i)
  \varphi(r_j')^*} {|\boldsymbol{r}_i-\boldsymbol{r}_j'|} ~\psi^*(\boldsymbol{r}_j')
\end{equation}
and similar formula is used for the $P$-wave interactions.
We find the mixed sequence $ S  $-wave  following $ P $-wave scattering terms small.
Hence, the summation via formula~(\ref{fm4}) is used separately for  $S~S$  and $P~P$  sequences.

Authors of Ref.~\cite{Kam01} find the necessity of an  additional term in the double scattering in $KD$  atoms because of $K^- \rightarrow K^0$ charge exchange. In the $1S$ atomic state it has large effect. The related charge exchange process in antiproton atoms corresponds to a process initiated with the first collision of an antiproton on a  proton which change into an
antineutron and a neutron. In a second collision on a neutron the antineutron changes back to an antiproton. The link of events is $$(  \bar{p}p )n        \rightarrow (\bar{n}n )n   \rightarrow  n(p\bar{p} )$$
and  corresponds to square of charge exchange amplitude  for $ \bar{p}p \rightarrow \bar{n}n $
given by isospin combination $ T_{ex} =( T_0-T_1)/2 $. On the other hand  $ \bar{p}p \rightarrow \bar{p}p $ amplitude is  given by $ T_{\bar{p}p}= (T_0+T_1)/2$.
In the scattering region the charge exchange cross section is an order of magnitude smaller then total cross $\bar{p}p$  section and $|T_{ex} /T_{\bar{p}p}|^2 \simeq 0.05$,  see  Ref. \cite{DED09}.
One could expect  the effect of the exchange process  to be  negligible but because of subthreshold $T_0$ dominance it turns to be  very small,  probably noticeable with improved potential models. One finds $|T_{ex} /T_{\bar{p}p}|^2 \approx 0.15 $ in the $S$ waves and $\approx 0.1$ in the $P$ waves  for both Paris models.
The real effect is additionally  reduced by statistical factors from numbers of available $(p,n) $ pairs versus all available nucleon pairs.
    As the exchange term follows $(T_0- T_1) ^2$  the  correction enter as small enhancement or small reduction of   the shifts and widths
    depending on positions of quasibound states. The calculated  numbers  have not been specified in the main text as the contributions are smaller then 1$\%$, much below the experimental precision.
On the other hand, if  the speculative Paris 09 suggestion  of the  80 MeV bound isospin~1  state turns out  to be true, the strong cancellations
    of $T_0 $ and $T_1$  amplitudes is removed in favour of $T_1$ dominance. The charge exchange scattering  may then matter.

    We are now in a position to check the convergence of expansion~(\ref{fm5})  in the  $2P$  atomic state of Helium. Forgetting the small exchange term in the triple scattering one obtains, with the Gaussian density profile,  $ C_1 = 0.133~ x$ and $C_2~=~8.92 \times 10^{-4} ~x^2$  with $x$ given by
    Eq.~(\ref{fm8}) where $N= A-1 $. The ratio $C_2/C_1 =  0.020~$x. Thus   $C_2$ terms contribute less than $1\%$ to the total shifts
    and are neglected.  On the other hand $C_1$ is to be calculated with the realistic nucleon wave function with the proper asymptotic behavior.

\end{document}